\documentclass[pra,twocolumn,showpacs,amsmath,amssymb]{revtex4}
\usepackage{epsfig}
\usepackage{color}
\usepackage{graphicx}
\begin{document}

\title{Ground state phase diagram of spin-$1/2$ bosons in
a two-dimensional optical lattice}
\author{L. de Forges de Parny$^1$, F. H\'ebert$^1$,
  V.G. Rousseau$^2$, R.T. Scalettar$^3$, and G.G. Batrouni$^{1,4}$}  
\affiliation{
$^1$INLN, Universit\'e de Nice-Sophia Antipolis, CNRS; 
1361 route des Lucioles, 06560 Valbonne, France,
}
\affiliation{
$^2$Department of Physics and Astronomy, Louisiana State University,
  Baton Rouge, Louisiana 70803, USA, 
}
\affiliation{
$^3$Physics Department, University of California, Davis, CA 95616,
}
\affiliation{
$^4$Centre for Quantum Technologies, National University of Singapore; 2
Science Drive 3 Singapore 117542.
}

\begin{abstract}
We study a two-species bosonic Hubbard model on a two-dimensional
square lattice by means of quantum Monte Carlo simulations.  In
addition to the usual contact repulsive interactions between the
particles, the Hamiltonian has an interconversion term which allows the
transformation of two particles from one species to the other. The phases
are characterized by their solid or superfluid properties and by their
polarization, {\it i.e.}  the difference in the populations. When
inter-species interactions are smaller than the intra-species ones,
the system is unpolarized, whereas in the opposite case the system is
unpolarized in even Mott insulator lobes and polarized in odd Mott
lobes and also in the superfluid phase.  We show that in the latter
case the transition between the Mott insulator of total density two
and the superfluid can be either of second or first order depending on
the relative values of the interactions, whereas the transitions are
continuous in all other cases.

\end{abstract}


\maketitle

\section{Introduction}

Since their first experimental realization, atomic Bose-Einstein
condensates have been used as a tool to study strongly interacting
quantum systems. This made possible the study of exotic phases,
especially in systems governed by Hamiltonians not easily realized in
condensed matter. Soon after the initial work on spin-0 bosons
\cite{bloch02}, much effort was directed at the problem of mixtures of
different kind of particles: fermions \cite{kett06} or mixtures of
bosons and fermions~\cite{kett07}. However, in these studies, the
effective spin degrees of freedom are frozen and the number of
particles of each type is generally fixed. Purely optical trapping
techniques allow to trap the atoms without freezing the internal
degrees of freedom. These systems exhibit both magnetic and superfluid
properties offering the opportunity to study the interplay between
these two effects. Unlike in simple mixtures, spin-spin interactions
are present in these systems and can be either ferromagnetic
($^{87}$Rb) or antiferromagnetic ($^{23}$Na)\cite{stamper,stamper2}.

Whereas current spin-$1$ experiments are in continuous space, a simple
model was proposed \cite{krutitsky04} for a system on an optical
lattice with only two internal low energy states.  The proposed system
consider neutral polarizable atoms with three degenerate atomic ground
states and three degenerate excited states characterized by the
magnetic quantum number $S_z = 0, \pm 1$.  In addition to generating a
periodic (optical lattice) potential, the counter propagating lasers
couple these internal ground and excited states by V and $\Lambda$
transitions which leads to only two low energy eigenstates denoted
respectively $0$ and $\Lambda$.  Such particles with only two internal
effective degrees of freedom are referred to as spin-1/2 bosons. As in
the spin-1 case, the interaction between these spins can be
ferromagnetic or antiferromagnetic and the presence of an optical
lattice allows the system to become strongly correlated with
superfluid (SF) or Mott insulator (MI) phases.  In previous work
\cite{deforges10}, we extended the original mean-field theory (MFT)
approach \cite{krutitsky04} and studied the one-dimensional system
with exact quantum Monte Carlo (QMC) simulations. Related spin-$1$
models were considered using MFT \cite{ashhab, kimura} or in one
dimension using QMC \cite{ggb09}. A similar model with two species of
bosons that can convert into each other was also studied
\cite{taka10}.

If the spin-spin interactions for the original spin one bosons are
ferromagnetic (F), the resulting on-site repulsions for spin 1/2
bosons are smaller between different particles than between identical
particles~\cite{deforges10,krutitsky04}.  In that case MFT and QMC
both show the system never polarizes \cite{deforges10}; in other
words, the populations of the two species always remain equal. If the
repulsion is strong enough, the system is in a Mott insulating phase
for commensurate densities and is otherwise superfluid. The
transitions between the superfluid and the Mott phases are always
continuous.

In the case of antiferromagnetic (AF) interactions for the original
spin one bosons, the repulsion between different particles is stronger
than the repulsion between identical particles
\cite{deforges10,krutitsky04} in the resulting spin 1/2 model and the
system exhibits a richer phase diagram. First, the Mott phases are
polarized for odd densities while they are unpolarized for even
ones. In addition, the superfluid phase is always polarized. Finally,
MFT predicts second order transitions between the odd density Mott
phases and the superfluid whereas the transitions between even MI and
SF can be first or second order depending on the strength of the
interaction. This was not observed in the one dimensional QMC
simulations \cite{deforges10}. This discrepancy was not surprising
since MFT often fails in reduced dimensionality and also since first
order phase transitions are generally absent in one dimension. Another
feature of the AF regime is a phase transition between a polarized and
an unpolarized phase inside the $\rho=1$ ($\rho$ being the total
particles density) Mott phase at low temperature observed with the QMC
simulations.

In this paper, we extend the QMC study of this system to the two
dimensional square lattice. The paper is organized as follows. In
section II, we will review the model and the techniques used to study
it. Sections III and IV will be devoted to the presentation of our
results for the cases of larger repulsion between different or
identical particles, respectively.  We summarize these results in
Section V.

\section{Spin-1/2 Model}

The model we will study has two species of bosonic particles governed
by the Hamiltonian \cite{krutitsky04},
\begin{eqnarray}
&&\mathcal{H} = -t \sum_{\sigma, \langle {\bf r,r'} \rangle}
\left(a^\dagger_{\sigma {\bf r}} a^{\phantom{\dagger}}_{\sigma {\bf r'}} +
a^\dagger_{\sigma {\bf r'}} a^{\phantom{\dagger}}_{\sigma {\bf r}} \right)
-\mu\sum_{\sigma, {\bf r}} \hat n_{\sigma {\bf r}} \label{h1}\\ && +\,
\frac{U_0}{2} \sum_{\sigma, {\bf r}} \hat{n}_{\sigma {\bf r}} (\hat
n_{\sigma {\bf r}} -1 ) + (U_0 + U_2) \sum_{\bf r} \hat{n}_{0 {\bf r}}
\hat{n}_{\Lambda {\bf r}} \label{h2}\\ && + \frac{U_2}{2} \sum_{\bf r}
\left( a_{0{\bf r}}^\dagger a_{0{\bf r}}^\dagger a_{\Lambda {\bf
r}}^{\phantom{\dagger}}
a_{\Lambda {\bf r}} ^{\phantom{\dagger}}+ a_{\Lambda {\bf r}}^\dagger a_{\Lambda {\bf
r}}^\dagger a_{0 {\bf r}}^{\phantom{\dagger}} a_{0 {\bf
r}}^{\phantom{\dagger}} \right)\label{h3},
\end{eqnarray}
where $a^\dagger_{\sigma {\bf r}}$ creates a particle of ``spin"
$\sigma = 0,\Lambda$ on site ${\bf r} = (x,y)$ of a periodic square
lattice of size $L\times L$.  The $\hat{n}_{\sigma {\bf r}} =
a^\dagger_{\sigma {\bf r}} a^{\phantom{\dagger}}_{\sigma {\bf r}}$ are
the corresponding number operators counting the particles of type
$\sigma$ on site $\bf r$.  The Hamiltonian includes the conventional
hopping term (Eq.~(\ref{h1})) which plays the role of a kinetic energy
term for lattice systems with the associated hopping parameter $t$
that is used as the energy scale.  To study the system in the grand
canonical ensemble, a chemical potential term is added to the
Hamiltonian (\ref{h1}). There are two density-density interaction
terms (Eq.~(\ref{h2})). The first describes repulsion between
identical particles on the same site with an associated energy $U_0 >
0$.  The second describes the on-site repulsion between particles of
different types. Depending on the value of $U_2$, this repulsion can
be stronger ($U_2>0$) or smaller ($U_2< 0$) than the repulsion between
identical particles. In this work, $|U_2|$ will remain smaller than
$U_0$ in order to keep only repulsive interactions. In most of our
work we kept $U_2/U_0$ fixed as $U_0$ is changed to map the phase
diagram.  Finally the last term of the Hamiltonian describes
conversion between the species: As two identical particles collide on
the same site, they can be converted into two particles of the other
kind. It was shown in Ref. \cite{krutitsky04} that the matrix element
for this conversion is equal to $U_2/2$, that is, it is essentially
the difference of interaction energies between identical and different
particles.  The sign of the conversion term (Eq.~(\ref{h3})) can be
chosen freely due to a symmetry of the model \cite{deforges10}.  Here
our convention has the opposite sign of the original
paper~\cite{krutitsky04}.

The only term in the Hamiltonian that couples different sites is the
hopping term in (\ref{h1}). MFT \cite{krutitsky04,deforges10} isolates
one site ${\bf r}$ and couples it to surrounding sites ${\bf r'}$ via
the mean values of destruction or creation operators $\psi_0 = \langle
a^{\phantom{\dagger}}_{0{\bf r'}}\rangle = \langle a^\dagger_{0{\bf
r'}}\rangle$ and $\psi_\Lambda = \langle
a^{\phantom{\dagger}}_{\Lambda {\bf r'}}\rangle = \langle
a^\dagger_{\Lambda {\bf r' }}\rangle$. This results in a one-site
Hamiltonian which can be easily diagonalized numerically. The ground
state energy is then minimized with respect to the two mean-field
order parameters $\psi_0$ and $\psi_\Lambda$. Normal or insulating
phases are obtained when $\psi_0 = \psi_\Lambda = 0$ whereas
superfluid phases occur whenever one of the $\psi$ is non zero.  To
study this system exactly, we used the stochastic Green function (SGF)
quantum Monte Carlo algorithm \cite{SGF, directedSGF}.  This algorithm
works in the canonical or grand canonical ensembles at finite inverse
temperature $\beta = 1/kT$. We generally used $\beta = 2L/t$ which is
usually a low enough temperature to obtain results that have converged
to their ground state limit for a system of linear size $L$ with
moderate interactions. However (see below) for the strongly
interacting regimes, it is sometimes necessary to use lower
temperatures (up to $\beta = 4L/t$).  In the canonical case with $N$
particles, the chemical potential is defined as the discrete
difference of the energy $\mu(N) = E(N+1) - E(N)$ which is valid in
the ground state where the free energy is equal to the internal
energy. The total particle density, $\rho$, is either fixed in the
canonical case or fluctuates in the grand canonical one. Densities for
particles of types 0 and $\Lambda$ are called $\rho_0$ and
$\rho_\Lambda$, respectively, and are not conserved due to the
conversion term Eq.~(\ref{h3}). The superfluid density is given by the
fluctuations of the winding number \cite{roy}
\begin{equation}
\rho_s = \frac{\langle (W_0 + W_\Lambda)^2\rangle}{4t\beta}
\label{rhos}.
\end{equation}
As explained in Ref.~\cite{roscilde10}, it is not meaningful to define
individual superfluid densities for the $0$ and $\Lambda$ particles as
their numbers are not conserved separately due to the conversion term
Eq.~(\ref{h3}). We also studied the single particle Green functions,
\begin{equation}
G_\sigma({\bf R}) =\frac{1}{2L^2}\sum_{\bf r} \langle a^\dagger_{\sigma {\bf
r+R}}a^{\phantom{\dagger}}_{\sigma {\bf r}} + a^\dagger_{\sigma {\bf
r}}a^{\phantom{\dagger}}_{\sigma {\bf
r+R}}\rangle,
\label{green1}
\end{equation}
where $\sigma = 0, \Lambda$. The Fourier transforms of the Green
functions, Eq.(\ref{green1}), give the momentum distributions,
$\rho_\sigma({\bf k})$.  The total density at zero momentum,
$\rho({\bf k}=0)$, is given by
\begin{equation}
\rho({\bf k}=0) = \sum_\sigma \rho_\sigma({\bf k}=0).
\label{totcond}
\end{equation}
Another useful quantity to characterize the Mott insulator is the
two-particle anti-correlated Green function $G_{\rm a}$
\begin{eqnarray}
G_{\rm a}({\bf R}) &=&\frac{1}{2L^2} \sum_{\bf r} \left\langle
a^\dagger_{\Lambda {\bf r}}a^\dagger_{0{\bf
r+R}}a^{\phantom{\dagger}}_{0{\bf r}}a^{\phantom{\dagger}}_{\Lambda
{\bf r+R}} + {\rm H.c.}  \right\rangle,
\label{greenf}
\end{eqnarray}
which is a measure of the exchange process whereby, for example, a $0$
particle is annihilated at site ${\bf r}$ and a $\Lambda$ particle at
site ${\bf r+R}$ while a $0$ particle is created at site ${\bf r+R}$
and a $\Lambda$ particle at site ${\bf r}$. If perfect phase coherence
is established by means of particle exchange, $G_{\rm a}$ reaches its
limiting upper value of $\rho_0 \rho_\Lambda$ at long distances ${\bf
R}$.

\section{Positive $U_2$ Case}

\begin{figure}
\includegraphics[width=8.5cm]{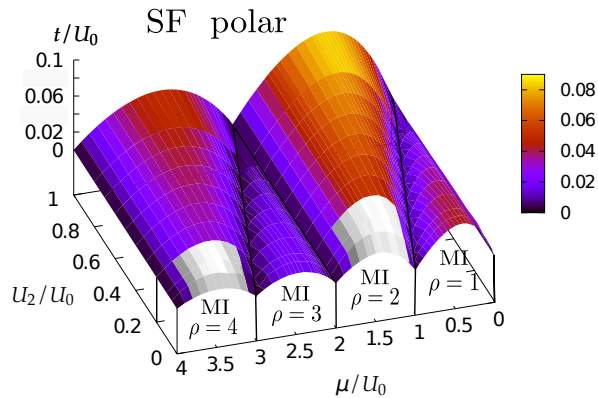}
\caption{(Color online) Zero temperature MFT phase diagram of the
bosonic spin-$1/2$ model in the $U_2>0$ case. The dome-shaped surface
shows the quantum phase transition between MI (below the surface) and
SF regions (above). The grey areas indicate the regions where the
transition is first order. Everywhere else the transitions are
continuous. The transitions are also second order for $U_2=0$.
\label{af_mf_diagram}}
\end{figure}

In the case where $U_2>0$, MFT predicts at zero temperature
\cite{deforges10} that the system is either in a polarized superfluid
(SF) phase or in incompressible Mott insulating (MI) phases. In the
polarized superfluid phase, the symmetry between the two populations
is broken and the density of one species becomes larger than the
other.  For total fillings which are odd multiples of the number of
sites, {\it i.e.} $\rho = 1, 3\cdots$, the MI phases are also
polarized whereas for even total densities, they are not:
$\rho_0=\rho_\Lambda$. The MI-SF transition is predicted to be
continuous for odd density Mott phases. For even densities, the
transition is found to be first order at the tip of the MI lobes for
small $U_2/U_0$ and becomes continuous as $U_2/U_0$ increases (see
Fig.~\ref{af_mf_diagram}). In the strongly interacting limit $t/U_0
\rightarrow 0$, the Mott phases of odd density $\rho$ are found for
chemical potentials $\mu$ in the interval $ \rho-1 < \mu/U_0 < \rho -
U_2 / U_0$, whereas the even density Mott phases are observed for
$\rho -1 -U_2 / U_0 < \mu/U_0 < \rho$. Hence as observed in
Fig.~\ref{af_mf_diagram} or Fig.~\ref{af_mf_qmc_diagram}, the odd
density Mott regions shrink as $U_2/U_0$ is increased and disappear
for $U_2/U_0 = 1$ in this $t=0$ limit.

Using QMC simulations in the one-dimensional case \cite{deforges10},
we observed the polarized SF phase, the polarized MI lobes for
$\rho=1$ and the unpolarized MI for $\rho=2$. However, contrary to MFT
predictions, all transitions were continuous; as is often the case in
one dimensional systems, no first order transitions were found.  We
also observed, at low temperature, that as $t/U_0$ decreases (always
keeping $U_2/U_0$ constant) the $\rho=1$ MI changes nature from
polarized to unpolarized.

\subsection{Phase diagram}

\begin{figure}
\includegraphics[width=8.5cm]{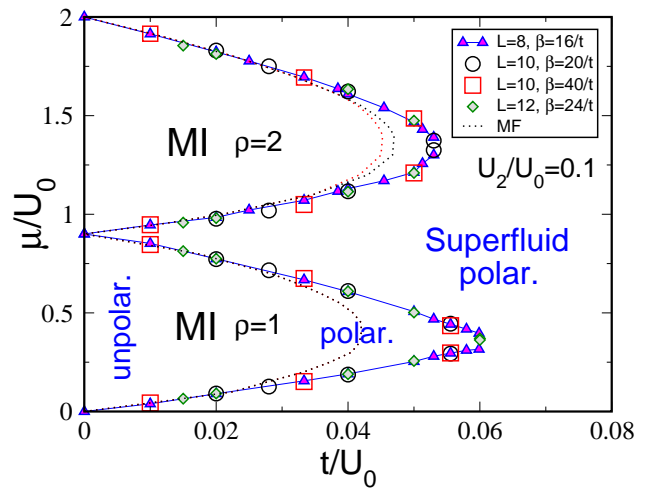}
\caption{(Color online) QMC phase diagram at $\beta=2L/t$ or $4L/t$
  for $U_2/U_0 = 0.1$ and several linear system sizes $L$. The
  superfluid phase is always polarised and the $\rho=2$ MI is not
  polarized due to the effect of the interspecies exchange term. For
  the $\rho=1$ MI phase, the system is polarized for the larger values
  of $t/U_0$ but then becomes unpolarized as $t/U_0$ decreases at
  finite $kT=t/2L$.  As $T$ decreases further to $kT=t/4L$
  polarization appears (see Fig.~\ref{histo_beta}).  The dotted lines
  are the mean-field results. The region between the two MFT curves in
  the $\rho=2$ lobe shows the coexistence zone due to the predicted
  first order transition.
\label{af_mf_qmc_diagram}}
\end{figure}

To map out the phase diagram, we employ the SGF algorithm in its
canonical implementation. The chemical potential is calculated using
energy differences to determine the boundaries of the MI lobes for
$\rho=1,2$. In addition, we studied the histograms of the densities of
the two species to determine whether a phase is polarized or not. We
found results similar to those found in one dimension: The superfluid
phase is always polarized, the $\rho=2$ MI phase is not polarized and
the $\rho=1$ shows a transition between a polarized MI and an
unpolarized MI as $t/U_0$ decreases. This point will be discussed in
more detail in Sec.~\ref{subpola}. The resulting phase diagram is
shown in Fig.~\ref{af_mf_qmc_diagram} for several linear lattice
sizes, $L$, and $\beta= 1/kT = 2L/t$ or $\beta=4L/t$.

The agreement between the QMC results and MFT is quite good. The Mott
lobes obtained with MFT are shown as dotted lines in
Fig.~\ref{af_mf_qmc_diagram} and are close to the exact boundary for
small values of $t/U_0$ with disagreement increasing as the tips of
the lobes are approached due to increased quantum fluctuations.  As
expected, the agreement is much better than in one dimension where we
had found a factor of two difference between the MFT predictions and
the observed lobe tips.

\begin{figure}
\includegraphics[width=8.5cm]{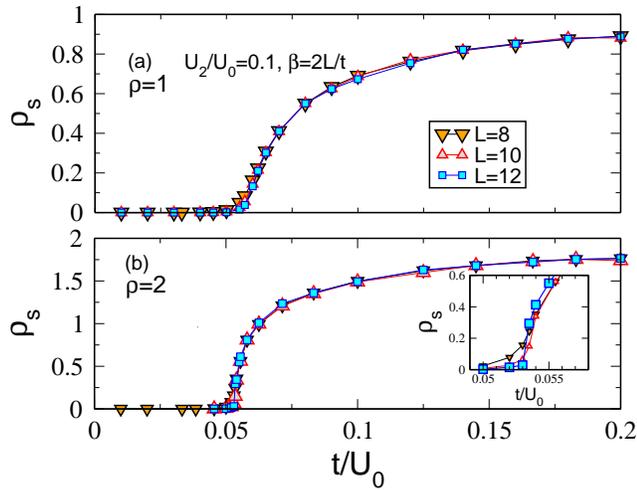}
\caption{The SF density, Eq. (\ref{rhos}) as a function of $t/U_0$ at
fixed total particle density for $U_2/U_0 = 0.1$. The transition from
the MI (small $t/U_0$) to the SF phase is signaled by $\rho_s$
becoming finite. For $\rho=1$ (a) the transition is continuous whereas
the jump in $\rho_s$ marks the presence of a first order phase
transition for $\rho=2$ (b). See Figs.~\ref{af_rho2_mf_qmc_rhos},
\ref{af_rho2_qmc_rho0}, and \ref{af_rho2_qmc_rho}. The inset shows the
transition region and the jumps in more
detail.\label{af_rho12_qmc_rhos}}
\end{figure}

\subsection{Phase transitions}

An important difference between one and two dimensions is that
first-order phase transitions may appear in the latter whereas they
are essentially excluded in the former. Since MFT predicts a first
order phase transition for even Mott lobes, we used QMC to study this
transition in great detail.

Figure \ref{af_rho12_qmc_rhos} shows the evolution of the superfluid
density $\rho_s$ at the MI-SF transition at fixed total density while
varying $t/U_0$, in other words the transition at the tip of the Mott
lobe. We observe that in the $\rho=1$ case
(Fig.~\ref{af_rho12_qmc_rhos}(a)), the transition is continuous. There
is no evidence at all for a first order transition for this case. On
the contrary, in the $\rho=2$ case (Fig.~\ref{af_rho12_qmc_rhos}(b)),
a jump in the superfluid density indicates a first order phase
transition.

\begin{figure}
\includegraphics[width=8.5cm]{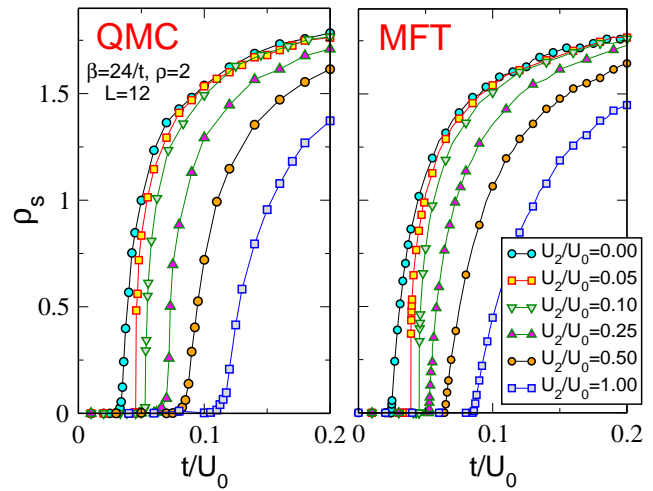}
\caption{(Color online) The superfluid density from QMC and MFT
calculations as a function of $t/ U_0$ for $\rho=2$ and for several
positive values of $U_2/U_0$. Both methods exhibit a jump in $\rho_s$
indicating the presence of a first order transition at the tip of the
$\rho=2$ Mott lobe for small but finite values of $U_2/U_0$. The jumps
increases from 0 at $U_2/U_0=0$ to a maximum for $U_2/U_0 \simeq 0.05$
and then decreases back to zero for $U_2/U_0 \ge 0.25$.
\label{af_rho2_mf_qmc_rhos} }
\end{figure}

\begin{figure}
\includegraphics[width=8.5cm]{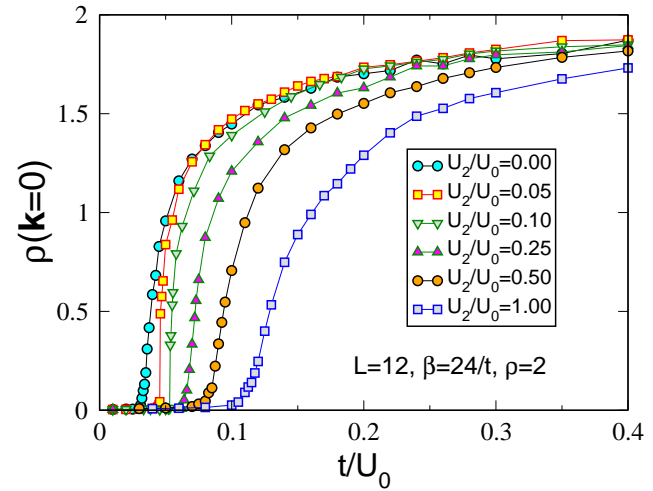}
\caption{(Color online) The density in the ${\bf k}=0$ momentum
$\rho({\bf k}=0)$ as a function of $t/ U_0$ for $\rho=2$ and for
several values of $U_2/U_0$. As for $\rho_s$,
Fig.~\ref{af_rho2_mf_qmc_rhos}, we observe a jump in $\rho({\bf k}=0)$
for small but finite values of $U_2/U_0$, confirming the presence of a
first order transition.
\label{af_rho2_qmc_rho0}}
\end{figure}

Comparing this $\rho=2$ case with MFT results for different values of
$U_2/U_0$ in Fig.~\ref{af_rho2_mf_qmc_rhos}, we observe very similar
behavior. The transition is continuous for $U_2/U_0=0$. Then it
becomes discontinuous for small values of $U_2/U_0 \lesssim 0.25$ and
is again continuous for $U_2/U_0 \ge 0.25$.  The jump in the
superfluid density varies continuously from zero at $U_2/U_0=0$ to a
maximum for $U_2 / U_0 \approx 0.05$ and then decreases back to zero
as $U_2/U_0$ is increased further. Figure \ref{af_rho2_qmc_rho0} shows
similar behavior for the particle density at zero momentum, $\rho({\bf
k}=0)$, where, once again, the jump at the transition is observed for
small values of $U_2/U_0$.

\begin{figure}
\includegraphics[width=8.5cm]{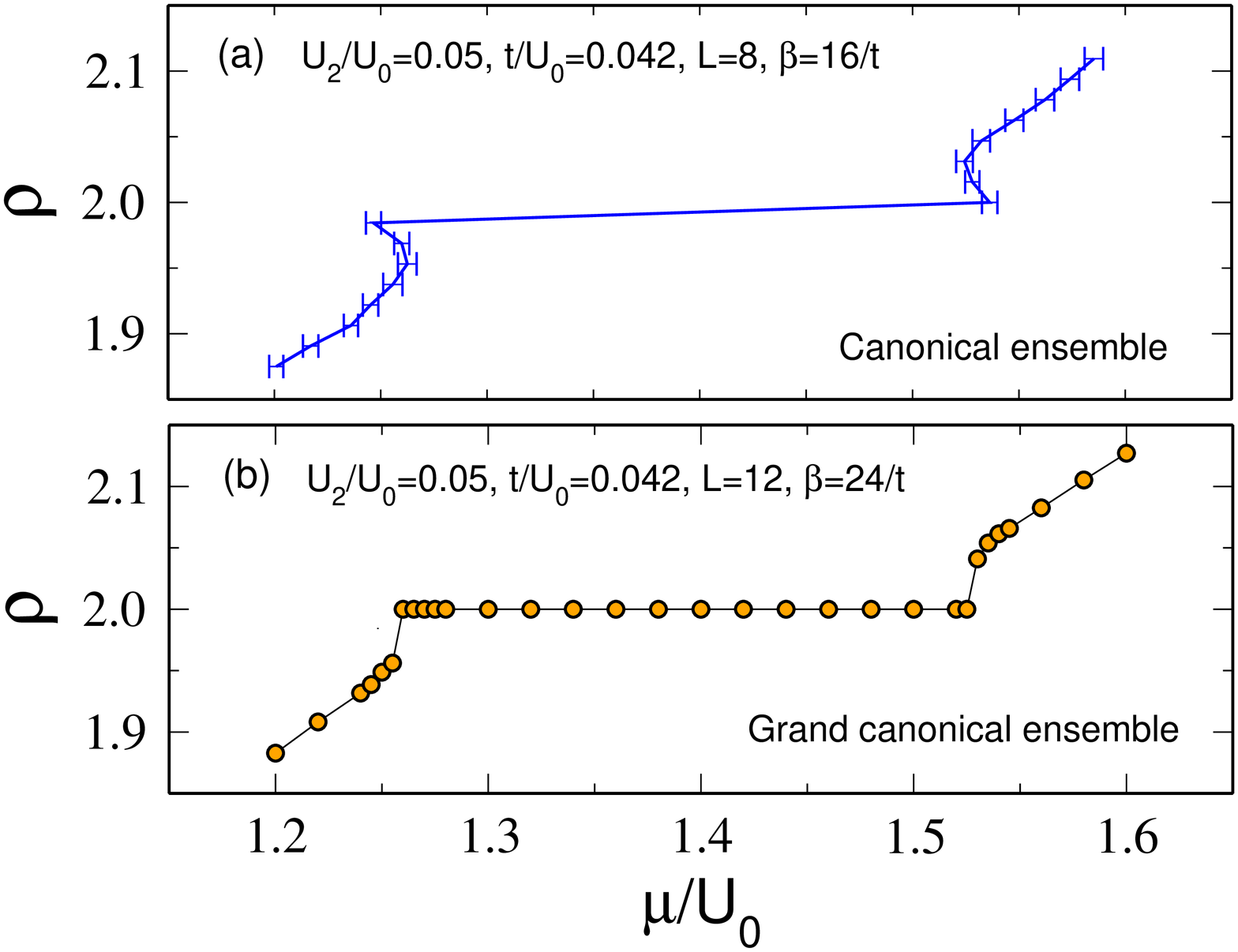}
\caption{(Color online) The density, $\rho$, as a function of the
chemical potential, $\mu$, close to the tip of the $\rho=2$ Mott
lobe. The canonical simulation (a) shows a negative compressibility
region, $\kappa\propto \partial \rho/\partial \mu < 0$. The grand
canonical simulation (b) exhibits a corresponding jump in the density.
Thus both approaches show the presence of a first order transition.
\label{af_rho2_qmc_rho}}
\end{figure}

To confirm the presence of first order phase transitions near the tip
of the MI lobe for even densities, we studied the behavior of $\rho$
as a function of $\mu$ as one cuts across the lobe at fixed $t/U_0$
using QMC simulation in the canonical and grand canonical
ensembles. In the canonical ensemble, a first order transition is
signaled by negative compressibility \cite{batrouniSS}, $\kappa
\propto \partial \rho/\partial \mu < 0$. In the grand canonical
ensemble, there will be a corresponding discontinuous jump in the
$\rho$ versus $\mu$ curve.  Figure \ref{af_rho2_qmc_rho} shows both
these cases. In Fig.~\ref{af_rho2_qmc_rho}(a) the canonical
simulations clearly show negative $\kappa$ just before and after the
Mott plateau at $\rho=2$.  On the other hand, the grand canonical
ensemble, Fig.~\ref{af_rho2_qmc_rho}(b), shows discontinuous jumps in
$\rho$ at the corresponding values of $\mu$. The canonical and grand
canonical simulations are in quantitative agreement on the size of the
unstable region which is extremely narrow; the system is stable for
densities smaller than $\rho=1.95$ or larger than $\rho = 2.05$ for
the chosen value of $t/U_0$.

\subsection{Polarization  and nature of the Mott phases\label{subpola}}

\begin{figure}
\includegraphics[width=8.5cm]{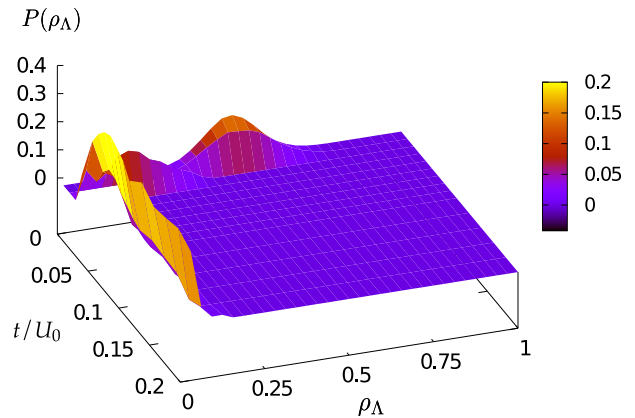}
\caption{(Color online) Histogram of the density of $\Lambda$
particles for $U_2 > 0$ as a function of $t/U_0$. The total density is
fixed at $\rho=1$ and the histogram of the $0$ particles is the image
of this distribution with respect to the line $\rho_\Lambda=0.5$. The
other parameters are $L=8$, $\beta = 2L/t$, and $U_2/U_0 = 0.1$. The
system polarizes for $t/U_0 \simeq 0.02$, well inside the $\rho=1$
Mott lobe. Thus at this low, albeit finite, temperature, there is a
transition between a polarized Mott phase at large $t/U_0$ and an
unpolarized Mott phase at lower $t/U_0$.  As shown in
Fig.~\ref{histo_beta} the complete lobe is polarized at $T=0$.
\label{af_histo_rho1}}
\end{figure}

We now analyse in more detail the polarization properties of the Mott
phases.  In the $\rho=1$ MI phase, Fig.~\ref{af_histo_rho1} shows that
a transition between unpolarized and polarized regions occurs for
$t/U_0 \simeq 0.02$ for $L=8$ and $\beta = 2L/t$.  This is well inside
the MI region as can be seen from the phase diagram,
Fig.~\ref{af_mf_qmc_diagram}. This possibility of a polarization
transition deep in the MI lobe requires closer examination.

In the $\rho=1$ MI phase, the system is frozen in a state with one
particle per site. Fluctuations around this state will occur when
particles hop around. However, events where two particles are
converted to the other species are negligible since this requires
double occupancies. Neglecting conversions and taking the hopping term
as perturbation to second order, the model can then be mapped onto an
anisotropic Heisenberg model \cite{kuklov03} where the presence of a
$\Lambda$ particle on a site corresponds to an up spin along the $z$
axis and a 0 particle corresponds to a down spin.  The interactions
between spins in the $xy$ plane and along the $z$ direction are given
by \cite{kuklov03}
\begin{eqnarray}
J_{xy} &=& -\frac{2t^2}{U_0}\frac{1}{1+U_2/U_0} \nonumber\\
J_z &=& -\frac{2t^2}{U_0}\frac{1+2U_2/U_0}{1+U_2/U_0}
\label{kuklov}
\end{eqnarray}
For $U_2>0$, we see that the couplings $J_{xy}$ and $J_{z}$ are always
negative, {\it i.e.} the effective model spin interactions are always
ferromagnetic. We also see that $|J_z| > |J_{xy}|$, which means that
ferromagnetic order will develop along the $z$ direction. In other
words, the system will become polarized in terms of 0 or $\Lambda$
particles in the ground state limit.

However, we also see that the energy associated with the polarization
of the system is $t^2 / U_0$ and becomes very small in the large $U_0$
limit. On the other hand, the histogram in Fig.~\ref{af_histo_rho1} is
at fixed temperature, $\beta=2L/t$, which means that the system is no
longer in the zero temperature limit for small values of $t/U_0 \le
0.02$ for $L=8$. Then excitations of these spin degrees of freedom
occur and the system is no longer polarized.

\begin{figure}
\includegraphics[width=8.5cm]{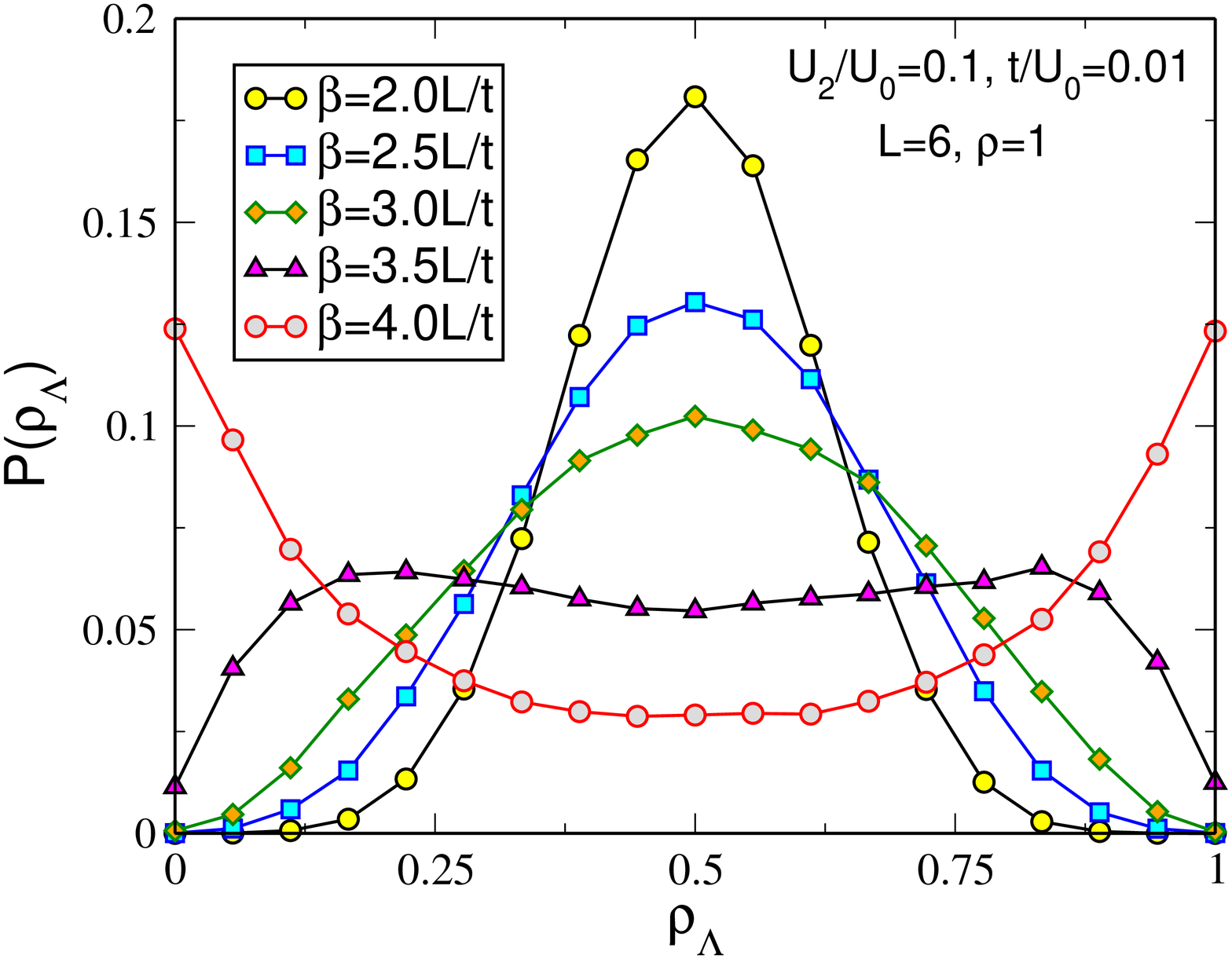}
\caption{(Color online) Histogram of the density of the $\Lambda$
particles as a function of temperature.  For large values of $U_0$,
one needs very small temperature for the system to polarize as the
energy associated to the polarization decreases when $U_0$ increases.
\label{histo_beta}}
\end{figure}

This argument is confirmed by QMC simulations done at lower
temperature: In Fig.~\ref{histo_beta} we show the evolution of the
histogram of $\rho_\Lambda$ for given values of $U_0$ and $U_2$ and
for several inverse temperatures from $\beta = 2L/t$ to $\beta =
4L/t$. We observe that as the temperature is decreased, the system
polarizes. We then conclude that, in the ground state limit, the
entire $\rho=1$ Mott phase is polarized, although one needs extremely
low temperatures in order to observe the polarization for large $U_0$.

\begin{figure}
\includegraphics[width=8.5cm]{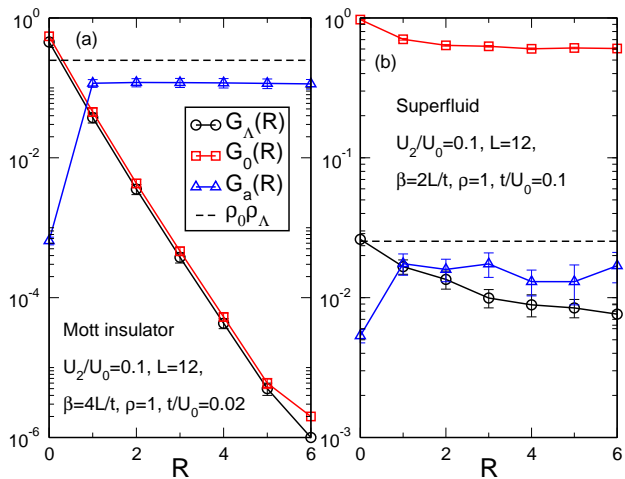}
\caption{(Color online) The single particle and the anticorrelated
Green functions for $U_2 > 0$ and $\rho=1$ along the $x$ axis in the
Mott insulator (a) and superfluid phases (b). In the MI phase,
individual movements of particles are strongly suppressed as shown by
the exponential decay of $G_0$ and $G_\Lambda$ whereas some
anticorrelated movements of particles remain, which is shown by the
plateau in $G_{\rm a}$.  In the superfluid phase, we find a long range
order for the Green function corresponding to the dominant species
$0$.
\label{af_green}}
\end{figure}

Figure \ref{af_green} shows single particle and anticorrelated Green
functions in the MI (a) and SF (b) phases. As expected, we see in the
Mott phase, Fig.~\ref{af_green}(a), that the individual Green
functions $G_0$ and $G_\Lambda$ decay exponentially to zero with
distance. The Heisenberg model approach predicts that, besides the
polarization of the system, ferromagnetic correlations in the $xy$
plane should be present. In the $xy$ plane spin-spin correlations are
measured with the correlation function $\langle S_{x,{\bf
r+R}}S_{x,{\bf r}} + S_{y,{\bf r+R}}S_{y,{\bf r}}\rangle$ which, in
terms of the particle creation and annihilation operators, maps into
the anticorrelated Green function $G_{\rm a}$, Eq. \ref{greenf}. In a
system where the density is fixed to one particle per site due to
interactions but where there are different species, there is always
the possibility that particles move by exchanging their
positions. $G_{\rm a}$ measures the coherence at long range of such
exchange moves. Indeed we see that such exchanges are present in the
$\rho=1$ Mott phase because $G_{\rm a}$ stays almost constant as
distance is increased, although it is smaller than its limiting value
$\rho_0 \rho_\Lambda$. This supports the description of the system, at
strong coupling, in terms of Heisenberg spins since individual
particle degrees of freedom appear to be irrelevant to the excitations
present in the system. On the other hand, spin excitations ({\it i.e.}
exchanges of particles) appear to be relevant.

In the superfluid phase, the Green function of the dominant species (0
in the case shown in Fig.~\ref{af_green} (b)) shows long range order,
indicating the presence of a long range phase coherence. We also
observe long range coherence for the minority species and for the
anticorrelated Green function. This is typical of a strongly
correlated superfluid where different kinds of phase coherence can be
observed: Phase coherence of the individual particles, but also, at
the same time, phase coherence of exchange moves of particles.

\begin{figure}
\includegraphics[width=8.5cm]{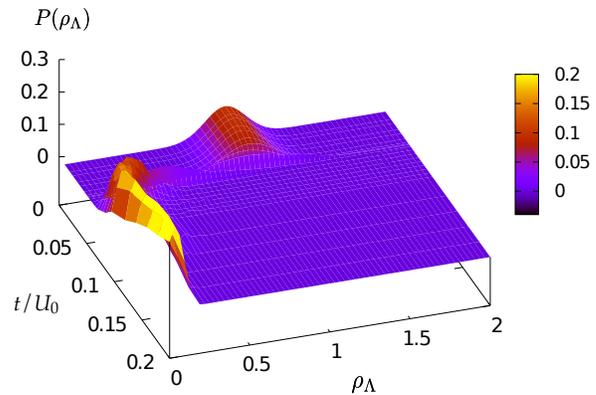}
\caption{(Color online) Histogram of the density of $\Lambda$
particles for $U_2 > 0$ as a function of $t/U_0$ at fixed total
density, $\rho=2$. The histogram of the $0$ particles is the image
of this distribution with respect to the line $\rho=1$. The other
parameters are $L=8$, $\beta = 2L/t$, $U_2/U_0= 0.1$. The
system polarizes for $t/U_0 \simeq 0.05$ which corresponds to the
MI-SF transition. Hence the $\rho=2$ Mott
is unpolarized.\label{af_histo_rho2}}
\end{figure}

In the $\rho=2$ MI case, the situation is simpler since the phase is
not polarized. With positive $U_2$, the system will adopt a state
where there are two particles of the same type on a given
site. However, the conversion term will couple this state with the
corresponding state with two particles of the other kind on the site,
thus lowering the energy. On average, those two states have the same
probability, and the mean number of a given type of particle on a site
is one. There is also no particular density ordering of the
system. This behavior is illustrated in Fig.~\ref{af_histo_rho2} where
the histogram of $\rho_\Lambda$ as a function of $t/U_0$ is shown. We
see that the transition between the unpolarized phase at low $t/U_0$
and the polarized phase happens for $t/U_0 \simeq 0.05$. Comparing
this with the phase diagram Fig.~\ref{af_mf_qmc_diagram}, we find that
the polarization occurs precisely at the MI-SF transition. Hence the
$\rho=2$ MI phase is not polarized but the superfluid phase is.

\begin{figure}
\includegraphics[width=8.5cm]{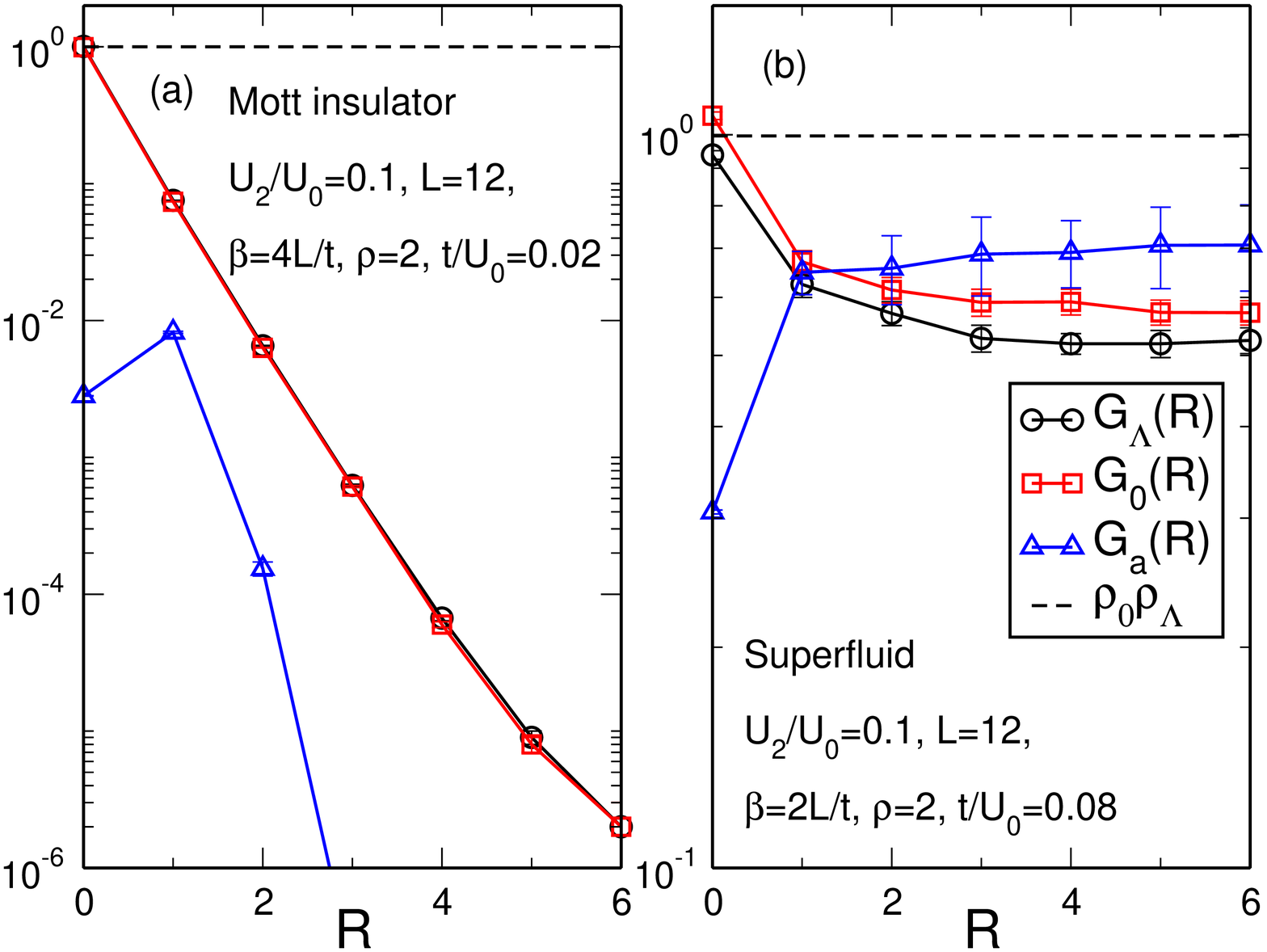}
\caption{(Color online) The single particle and the anticorrelated
Green functions for $U_2>0$ and $\rho=2$ along the $x$ axis in the
Mott (a) and superfluid phases (b). In the MI phase, individual
movements as well as exchanges of particles are suppressed, which is
shown by the exponential decay of all the Green functions. Phase
coherence is established in the SF phase (b).
\label{af_green_rho2}}
\end{figure}

Neglecting the hopping term, there is no degeneracy in the state
adopted by the system in the $\rho=2$ Mott phase. Then the
fluctuations induced by the hopping term do not create any phase
coherence and all the Green functions decay exponentially
(Fig.~\ref{af_green_rho2}(a)). In the superfluid phase
(Fig.~\ref{af_green_rho2}(b)), we observe a behavior similar to the
$\rho=1$ case where all the Green functions reach a plateau at large
distance, thus showing the presence of different kinds of phase
coherence in the system.  Due to the larger density and the importance
of conversion processes in this case, the polarization of the system
is weaker than in the $\rho=1$ case and the leading phenomenon seems
to be exchanges as $G_{\rm a}$ is larger than $G_0$ or $G_\Lambda$.

\section{Negative $U_2$ case}

\begin{figure}
\includegraphics[width=8.5cm]{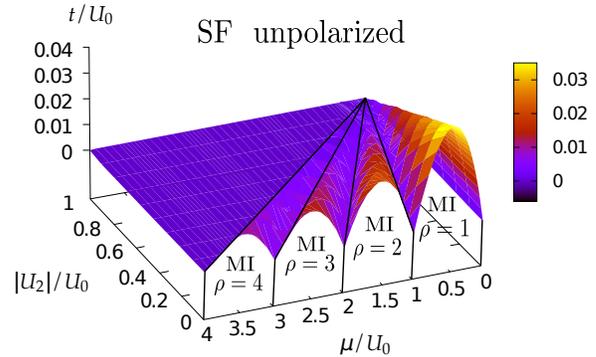}
\caption{(Color online) Ground state phase diagram of the $U_2 < 0$
case in MFT for $\rho \le 4$. The surface traces the transition points
between MI phases (below the surface) and SF regions (above). As
$|U_2|/U_0\to 1$, the MI regions shrink and eventually disappear.
\label{mf_diagram_F}}
\end{figure}

In the negative $U_2$ case, the phase diagram predicted by MFT, and
observed in one-dimensional QMC simulations is less rich than its
positive counterpart.  As discussed in Section II, we focus on the
range of interactions $|U_2|/U_0 < 1$. Figure \ref{mf_diagram_F} shows
the MFT phase diagram which exhibits MI and SF phases, both of which
are unpolarized. As $U_2/U_0 \to -1$ and for $t/U_0 = 0$, all the MI
lobes, with even or odd densities, shrink and totally disappear at
$U_2/U_0 = -1$ as the MI phase of density $\rho$ is obtained for
$(1+U_2/U_0)(\rho-1) < \mu/U_0 < (1+U_2/U_0)\rho$.  MFT predicts that
all phases are unpolarized and all transitions continuous. This was
confirmed in one dimension with QMC simulations.  Here, we focus on
the two dimensional case.

\subsection{Phase diagram and transitions}

\begin{figure}
\includegraphics[width=8.5cm]{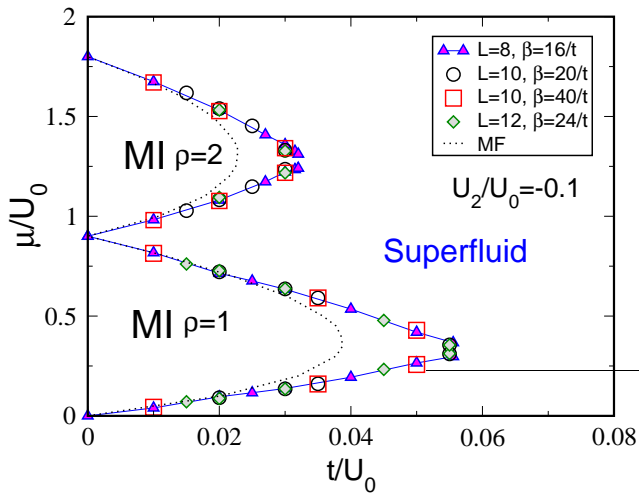}
\caption{(Color online) QMC phase diagram at low temperature for three
system sizes in the $U_2 < 0$ case. Unlike the $U_2 > 0$ case, the
system never polarizes. The dotted lines show the MFT
result.\label{f_phase_diagram}}
\end{figure}

Fig.~\ref{f_phase_diagram} shows the QMC phase diagram for the
negative $U_2$ case at $U_2/U_0 = -0.1$. The boundaries of the MI
lobes are obtained in the same way as in the positive $U_2$ case.  The
results are qualitatively similar to those found in one dimension. The
system exhibits MI lobes at commensurate fillings and sufficiently
small $t/U_0$ which turn superfluid as this parameter increases. In
addition the system is SF for all incommensurate fillings.  We find
that both these phases are always unpolarized (see below).  As for
$U_2 > 0$, the agreement between QMC and MFT is better in two
dimensions than in one, especially at small $t/U_0$. The agreement is
poor as the tips of the lobes are approached due to increased quantum
fluctuations.

\begin{figure}
\includegraphics[width=8.5cm]{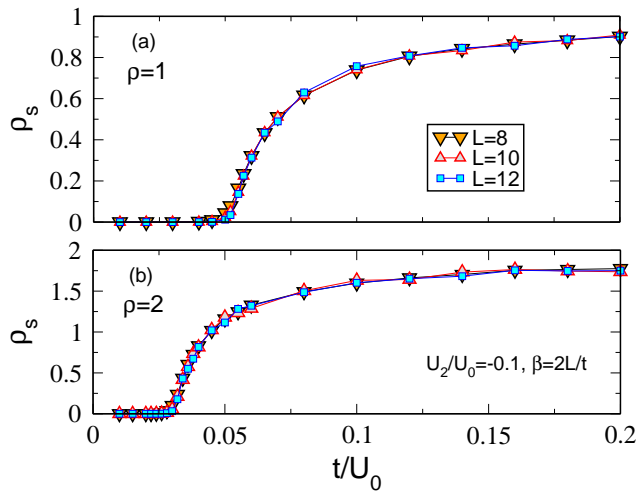}
\caption{(Color online) The SF density, $\rho_s$, as a function of
$t/U_0$ at fixed density for $U_2 < 0$. The superfluid density varies
continuously for $\rho=1$ (a) and $\rho=2$ (b), showing no sign of a
first order phase transition.
\label{f_transition_rho12}}
\end{figure}

In contrast with the $U_2 > 0$ case, all quantum phase transitions
appear to be continuous in this case. As can be observed in
Fig.~\ref{f_transition_rho12}, there are no signs of possible
discontinuities in the superfluid density at the transition between
the MI and SF phases. Other quantities, such as $\rho({\bf k}=0)$ (not
shown here) confirm this conclusion.

\subsection{Polarization and nature of the Mott phases}

\begin{figure}
\includegraphics[width=8.5cm]{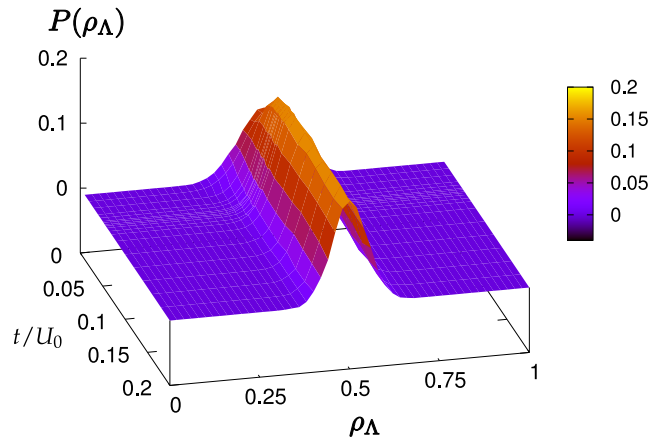}
\caption{(Color online) Histogram of the density for particles of type
$\Lambda$ for $U_2 > 0$ for a total density $\rho=1$ and $U_2/U_0 =
-0.1$, $L=10$ and $\beta=2L/t$.  The histogram is always centered
around $\rho_0 = \rho_\Lambda = 0.5$ and the system never becomes
polarized in this case. There is no change of behavior when the MI-SF
transition is crossed for $t/U_0 \simeq 0.05$.
\label{f_histo_rho1}}
\end{figure}

To analyse the situation in the $\rho=1$ MI, we again use the
perturbative mapping to the anisotropic Heisenberg model
(Eq. \ref{kuklov}).  We restrict ourselves to the case where all the
interactions are repulsive : $-1< U_2/U_0 < 0$.  In this range
$|J_{xy}| > |J_z|$ and the system develops spin ordering in the $xy$
plane.  In terms of the original model representing two types of
particles, this spin order in the $xy$ plane corresponds to a phase
with equal number of $0$ and $\Lambda$ particles on each site.  This
is confirmed by QMC simulations. Plotting the density histograms for
$U_2/U_0 = -0.1$, Fig.~\ref{f_histo_rho1}, it is seen that the
distribution is always centered around $\rho_0 = \rho_\Lambda = 0.5$
confirming that the system is not polarized. In addition, the
density-density correlations (not shown) do not exhibit any sign of
density order.

\begin{figure}
\includegraphics[width=8.5cm]{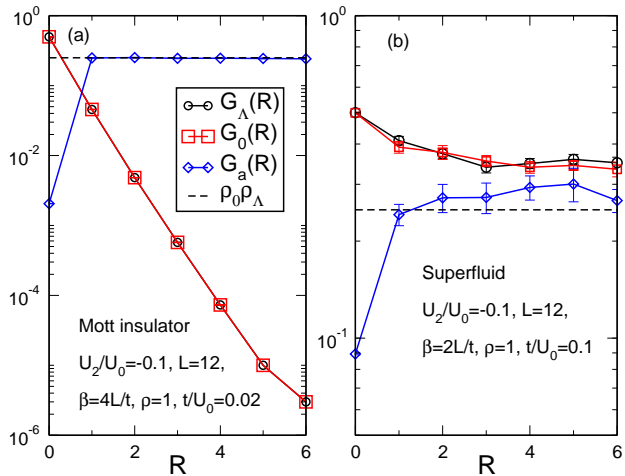}
\caption{(Color online) The single particle and anticorrelated Green
functions for $U_2 > 0$ and $\rho=1$ along the $x$ axis in the MI (a)
and SF phases (b). For the MI phase, as predicted by the spin
approximation, we find a dominant $G_{\rm a}$ showing that the
displacement of particles is due mainly to exchange of particles of
different types whereas movements of individual particles are
suppressed. For the superfluid phase, there is long range coherence of
all the Green functions.
\label{f_green}}
\end{figure}

In this case, the prediction of the mapping onto the Heisenberg model
is that the system should develop ferromagnetic order in the $xy$
plane and the spin-spin correlations $\langle S_{x,{\bf r+R}}S_{x,{\bf
r}} + S_{y,{\bf r+R}}S_{y,{\bf r}}\rangle$ should be long
ranged. Figure \ref{f_green} shows the single particle and the
anticorrelated Green functions in the MI (a) and the SF (b) phases. In
the $\rho=1$ MI phase, Fig.~\ref{f_green}(a), we see that $G_0({\bf
R})$ and $G_\Lambda({\bf R})$ decay exponentially as expected.
However, we also see that $G_{\rm a}({\bf R})$ quickly saturates to a
constant value at large separations indicating that exchange moves are
common.  In a system where the number of particles of each species is
fixed, {\it i.e.} no conversion between species, this behavior of
$G_{\rm a}({\bf R})$ would correspond to a counter superfluid phase
(CSF) \cite{kuklov03}. However, in this system, due to the residual
effects of the conversion term, it is not possible to calculate a
corresponding counter superfluid density \cite{roscilde10}. However,
we do observe the presence of a long range coherence of exchange moves
as predicted by the perturbative mapping to the Heisenberg spin
system.  The plateau observed in $G_{\rm a}$ takes on its maximum
possible value at large distances $G_{\rm a} \rightarrow \rho_0
\rho_\Lambda$, showing perfect phase coherence, contrary to what was
observed for $U_2 > 0$.

In the superfluid phase at $\rho=1$, Fig.~\ref{f_green} (b), $G_0$ and
$G_\Lambda$ exhibit long range order, indicating the presence of phase
coherence. The two species remain correlated, of course, since the
system is still in a strongly interacting regime ($t/U_0= 0.1$).  This
correlation can be observed, for example, from the fact that $G_{\rm
a}$ is larger than the product $G_0 G_\Lambda$. This means that, while
particles can move independently, exchanges of different particles are
still present.

\begin{figure}
\includegraphics[width=8.5cm]{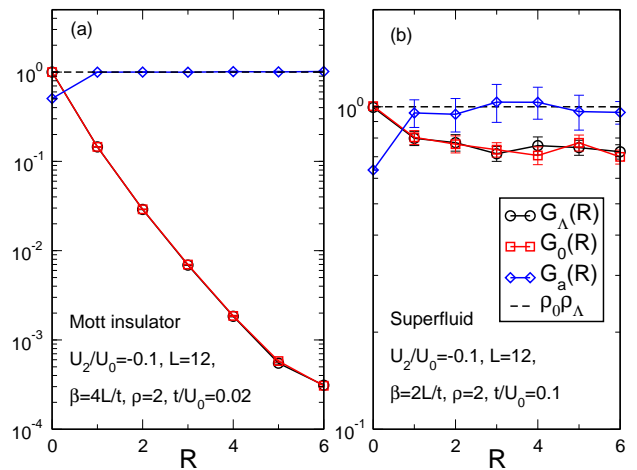}
\caption{(Color online) The single particle and anticorrelated Green
functions for $U_2 < 0$ and $\rho=2$ along the $x$ axis in the MI (a)
and SF phases (b).  For the MI phase, there are exchanges of particles
due to the degeneracy of the Mott phase in the $t=0$ limit, whereas
individual movements are prohibited. In the SF, all Green functions
show phase coherence of individual movements of particles or of
anticorrelated movements.
\label{f_green_rho2}}
\end{figure}

The $\rho=2$ Mott phase is once again unpolarized (the histogram is
similar to the $\rho=1$ histogram shown in
Fig. \ref{f_histo_rho1}). In the $t/U_0 \rightarrow 0$ limit, there
exist two possible degenerate ground states on each site.  The first
state is obtained by putting one particle of each type on a given
site, thus obtaining an interaction energy $U_0 + U_2 < U_0$ as $U_2 <
0$. The second state is the superposition of a state with two 0
particles and a state with two $\Lambda$ particles: $(|00\rangle +
|\Lambda \Lambda\rangle) / \sqrt{2}$.  While the interaction terms
Eq. \ref{h2} gives in this case an energy $U_0$, it is reduced to $U_0
+ U_2$ by the conversion term Eq. \ref{h3}.  For these two possible
states, the mean densities $\rho_0$ and $\rho_\Lambda$ are equal and
the system does not polarize which is directly observed in density
histograms (not shown here).  The $\rho=2$ Mott ground state then
shows a large degeneracy in the $t=0$ on-site limit. All these
degenerate states are once again coupled by second order contributions
from the hopping term and the degeneracy is lifted by establishing a
phase coherence of exchange movements. This can be understood in the
following way: whether a site is occupied by two 0 particles, two
$\Lambda$ particles or one 0 and one $\Lambda$ particle, the on site
energy is the same, because the interaction energy is lower for
different particles or because it is lowered by conversion term for
identical particles. Then starting from any configuration, exchanges
of particles of different types will exchange the states of
neighboring sites without changing the on site energy.  A phase
coherence is then established by exchange moves in the Mott phase,
reminiscent of the one observed in the $\rho=1$ case.  This is shown
in Fig.~\ref{f_green_rho2}(a) where we show perfect phase coherence of
the $G_{\rm a}$ function which reaches its limiting value of $\rho_0
\rho_\Lambda = 1$ in the Mott phase whereas individual Green functions
decay exponentially.

In the superfluid phase, we observe phase coherence both for exchange
moves and for individual movements of particles but, once again, the
exchange moves appear to give the leading contribution
(Fig.~\ref{f_green_rho2}(b)).

\section{Conclusion}

In this work, we used QMC simulations to determine the phase diagram
of the spin-1/2 bosonic Hubbard model on a two-dimensional square
lattice.  For $U_2 > 0$, we found that the numerical results to be in
good agreement with MFT, especially at very small $t/U_0$ where
quantum fluctuations are highly suppressed.  The system has three
phases: A polarized superfluid, an unpolarized Mott phase for $\rho=2$
(and likely for all even MI lobes) and a polarized Mott phase for
$\rho=1$ (and likely for all odd MI lobes). We have shown that the
polarization of this $\rho=1$ Mott phase can be understood in terms of
an effective anisotropic Heisenberg model and that, in the ground
state, the phase is completely polarized, although extremely low
temperatures are needed to observe the polarization in the strongly
interacting regime.  The $\rho=2$ phase is not polarized due to the
action of the conversion term that will transform pairs of identical
particles into pairs of the other types of particles and then suppress
possible polarization of the system. The first order MI-SF transition
for the $\rho=2$ MI phase which was predicted by MFT is confirmed
numerically.  In addition, the QMC results showed that in the $\rho=1$
MI lobe, moves which exchange the positions of a $0$ and a $\Lambda$
particles are present as evidenced by the saturation of the
anticorrelated Green function $G_{\rm a}$.  This property was not
previously addressed by MFT calculations.

For $U_2 < 0$, the system is always unpolarized and MI-SF transitions
are all continuous. However, in the MI phases, the origin of this
absence of polarization is different for the $\rho=1$ and the $\rho=2$
phases. For $\rho=1$, the anisotropic Heisenberg model approach shows
the domination of an effective in-plane coupling, whereas for
$\rho=2$, the two possible degenerate states on each site are both
unpolarized. Here too, we found that particle exchange moves are
present in the $\rho=1$ and, more surprisingly, in the $\rho=2$ Mott
insulator.

\acknowledgements

This work was supported by: the CNRS-UC Davis EPOCAL LIA joint research
grant; by NSF grant OISE-0952300; an ARO Award W911NF0710576 with funds
from the DARPA OLE Program.

\end{document}